\documentclass[%
 reprint,
superscriptaddress,
 amsmath,amssymb,
 aps,
pra,
]{revtex4-2}

\usepackage{graphicx}
\usepackage{dcolumn}
\usepackage{bm}
\usepackage{braket}
\usepackage{subcaption}
\usepackage{mathtools}
\usepackage{float}

 \usepackage{todonotes} 

\begin{document}

\title{Efficient Adiabatic Rapid Passage in the Presence of Noise}

\author{Kehui Li}
\affiliation{Centre for Quantum Information and Quantum Control, Department of Physics, University of Toronto, Toronto, Ontario, Canada.}
\author{David C. Spierings}
\affiliation{Centre for Quantum Information and Quantum Control, Department of Physics, University of Toronto, Toronto, Ontario, Canada.}
\author{Aephraim M. Steinberg}
\affiliation{Centre for Quantum Information and Quantum Control, Department of Physics, University of Toronto, Toronto, Ontario, Canada.}
\affiliation{Canadian Institute For Advanced Research, Toronto, Ontario, Canada.}

\date{\today}

\begin{abstract}
Adiabatic Rapid Passage (ARP) is a powerful
technique for efficient transfer of population between quantum states.
In the lab, the efficiency of ARP is often limited
by noise on either the energies of the states or the frequency of the driving field. We study ARP in the simple setting of a two-level system subject to sinusoidal fluctuations on the energy level separation by numerically solving the optical Bloch equations in the absence of damping. We investigate the dependence of the efficiency of population transfer on the frequency and amplitude of the perturbation, and find that it is predominantly affected by resonant coupling when the detuning matches the frequency of the noise.  We present intuitive principles for when ARP becomes inefficient within this model, and provide a sufficient condition for the population transfer to be above an arbitrary threshold.
\end{abstract}

\maketitle

\section{Introduction}
Coherent population transfer is an important tool for state preparation in many systems, such as trapped ions \cite{trappedIons}, Rydberg atoms \cite{rydberg1,rydberg2}, molecules \cite{molecule2,molecule1} and more. A common technique to achieve efficient population transfer in a two-level system is Adiabatic Rapid Passage (ARP), which typically sweeps the detuning of a driving field  across resonance slowly in order to achieve full population transfer. Compared to other population inversion methods such as a $\pi$-pulse (i.e.~a resonant pulse with time-integrated area of $\pi$), an advantage of ARP is that it does not depend on precise control of the resonant frequency, or the amplitude and timing of the pulse. Therefore, ARP is robust against small changes in resonance, making it especially effective for population inversion in inhomogeneously broadened samples.

\medskip
For ARP to be successful, the frequency sweep must be both ``rapid" and ``adiabatic", as its name suggests. The sweep should be fast compared to the excited state lifetime, while slow compared to the period of a resonant Rabi oscillation.
By the adiabatic theorem, if the sweep is sufficiently slow, then the quantum state of the system adiabatically follows the instantaneous energy eigenstate \cite{adiabaticThm}. 
Far from resonance, the bare states and those dressed by the driving field coincide, while at resonance an avoided-crossing occurs. The lower eigenstate is adiabatically connected to the ground state when detuning $\delta \to -\infty$, and is connected to the excited state when $\delta \to +\infty$. Therefore, when an atom starts in the ground state and the field detuning $\delta$ slowly changes from $-\Delta$ to $\Delta$ (where $\Delta \to \infty$), the atom remains in the lower eigenstate, becoming excited by the end of the chirp, thus achieving population inversion. In the case where detuning is chirped linearly in time, the efficiency of population transfer by ARP is given by the Landau-Zener formula \cite{LZFormula}, and depends on the speed of the sweep.
\medskip

In practice, ARP is performed in finite time, over a finite frequency range, and is subject to noise from the environment. Therefore, studying the performance of ARP in these realistic scenarios is crucial to testing its robustness. There have been many studies on the effect of both ``off-diagonal noise," which is related to fluctuations of the coupling strength between levels, and ``diagonal noise," which is related to fluctuations of the separation between energy levels. Off-diagonal coloured Gaussian noise has been studied in \cite{fastColouredGaussian, colouredGaussian}, slow off-diagonal noise in \cite{LZslowNoise}, and off-diagonal harmonic noise in \cite{off-diag}. Here, we investigate the effect of diagonal noise, which has been studied in the large amplitude and rapid fluctuation limit with a stochastic model in \cite{stocastic0,stochastic1}.
We aim to understand the effect of perturbations with arbitrary profiles, by focusing on the sinusoidal time-variation of the energy level separation due to a single Fourier component of the noise. The setup of the problem and the system Hamiltonian will be presented in Section~\ref{section2}. The results of our simulations and their implications will be discussed in Section~\ref{section3}. In particular, we investigate the role of resonances between the noise and the system, and study how the effect depends on the frequency and amplitude of the noise. We also present a comparison between our simulations and a non-adiabatic model presented in \cite{model} that noticed similar behavior. Finally, we end with a discussion on noise tolerance, and find a condition for good population transfer despite the noise.

\section{The Two-Level System and Hamiltonian} \label{section2}
\begin{figure}[h]
\centering
\includegraphics[width=0.7\linewidth]{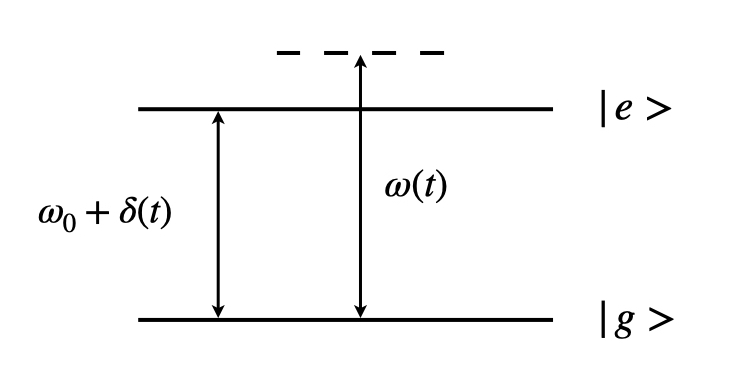}
\caption{Energy level diagram}
\label{fig:energy level}
\end{figure}

We consider a two-level system driven by an external field with linearly chirped frequency. The energy difference between the levels is subject to a sinusoidal perturbation, which represents the noise on the system. An energy level diagram is shown in Figure~\ref{fig:energy level},
where $\hbar \omega_0$ is the unperturbed energy difference between the two levels, and $\delta(t)$ is the perturbation, given by $\delta(t)=\delta_{osc}\cos(\omega_{osc}t+\phi)$.
The frequency of the driving field is chirped linearly across resonance so that $\omega(t) = \omega_0 + \dot{\omega}t$. The detuning between the driving field and the system is therefore $\Delta(t) = \dot{\omega}t - \delta_{osc}\cos(\omega_{osc}t + \phi)$.

\medskip
Under the Rotating-Wave Approximation (RWA), the Hamiltonian of the system can be written as 
\begin{equation}
    \hat{H}(t)=\hbar \left( \frac{\Omega_0}{2}\ket{e}\bra{g} + \frac{\Omega_0}{2}\ket{g}\bra{e} - \Delta(t)\ket{e}\bra{e} \right),
\end{equation}
where $\ket{g}$, $\ket{e}$ are the ground and excited states and $\Omega_0$ is the bare Rabi frequency. For simplicity, we set $\hbar=1$ in the following discussions.

\medskip

\begin{figure}[h]
\centering
\subcaptionbox{ \label{fig:small amp}}[.23\textwidth]{
  \includegraphics[width=1\linewidth]{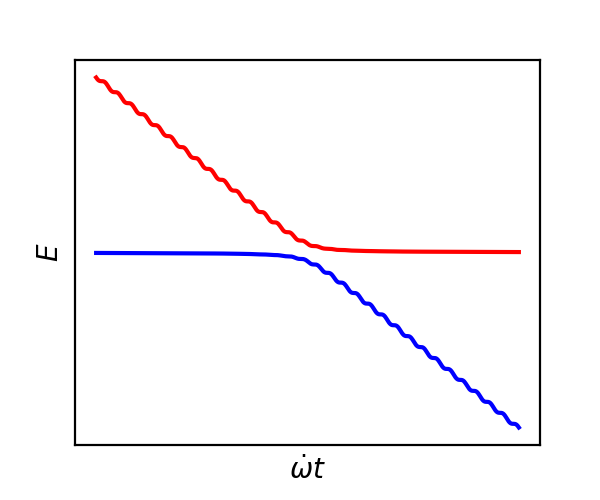}}
  \hfill
\subcaptionbox{ \label{fig:small freq}}[.23\textwidth]{
  \includegraphics[width=1\linewidth]{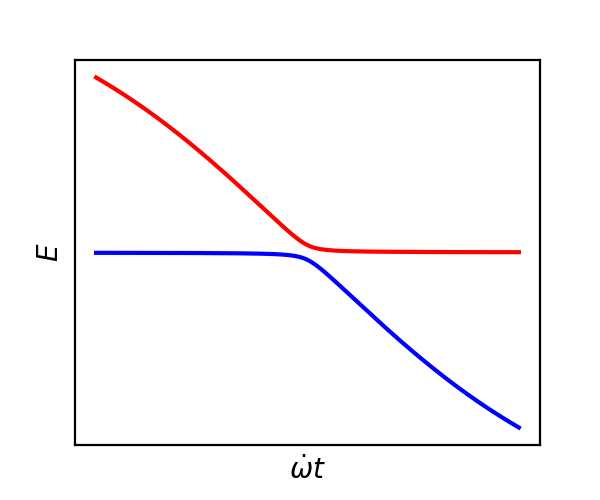}}

\subcaptionbox{ \label{fig:else}}[.23\textwidth]{
  \includegraphics[width=1\linewidth]{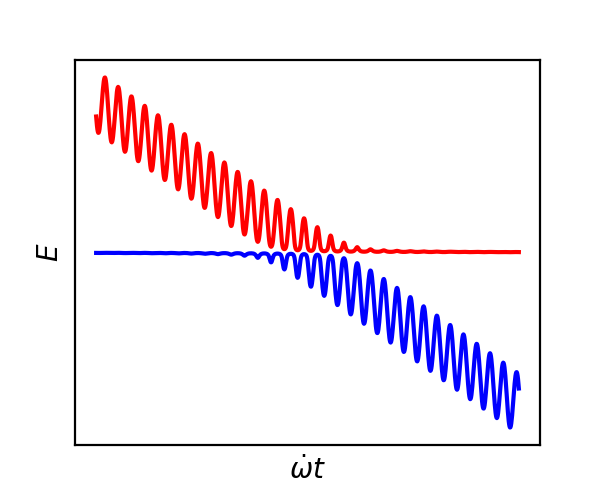}}
  
\caption{Instantaneous eigenenergies of the Hamiltonian in three regimes. (a) shows a noise with a small amplitude, which can be treated as a small perturbation to the linear sweep. (b) shows a slow noise, which can be considered a linear sweep with a modified slope and shifted endpoints. The scenario in (c), with a moderate noise frequency and amplitude, is the focus of this paper.}
\label{fig:dressed states}
\end{figure}

Figure \ref{fig:dressed states} shows the instantaneous eigenvalues of the Hamiltonian in three different regimes. In the limit of small-amplitude noise (i.e.~$\delta_{osc} \ll \Omega_0$, Fig \ref{fig:small amp}), the noise is a weak perturbation on the Landau-Zener Hamiltonian, $\hat{H}_0(t)= \frac{\Omega_0}{2}\ket{e}\bra{g} + \frac{\Omega_0}{2}\ket{g}\bra{e} - \dot{\omega}t \ket{e}\bra{e}$, and one would expect the effect of the noise on the population transfer to be very weak. In the limit of small noise frequency (i.e.~$2\pi/\omega_{osc} \gg $ duration of the sweep, Figure \ref{fig:small freq}), the noise appears almost linear during the sweep time, $\delta(t)\approx \delta_{osc}\cos(\phi) - \omega_{osc} \delta_{osc} \sin(\phi) t$, so that the detuning as a function of time can be viewed simply as a new linear sweep with slope $\dot{\omega}- \omega_{osc} \delta_{osc} \sin(\phi)$ and shifted endpoints. When $2\pi/\omega_{osc}$ is much less than the duration of the sweep and the amplitude of the noise is moderate, as shown in Figure \ref{fig:else}, the effect of the noise can no longer be explained by the simple intuitions for the previous two limiting cases. This regime is the focus of discussion in the following section.

\section{Discussion}\label{section3}
\subsection{Noise resonances}
\begin{figure*}[p!]
\centering
\begin{subfigure}{.33\textwidth}
  \centering
  \includegraphics[width=1\linewidth]{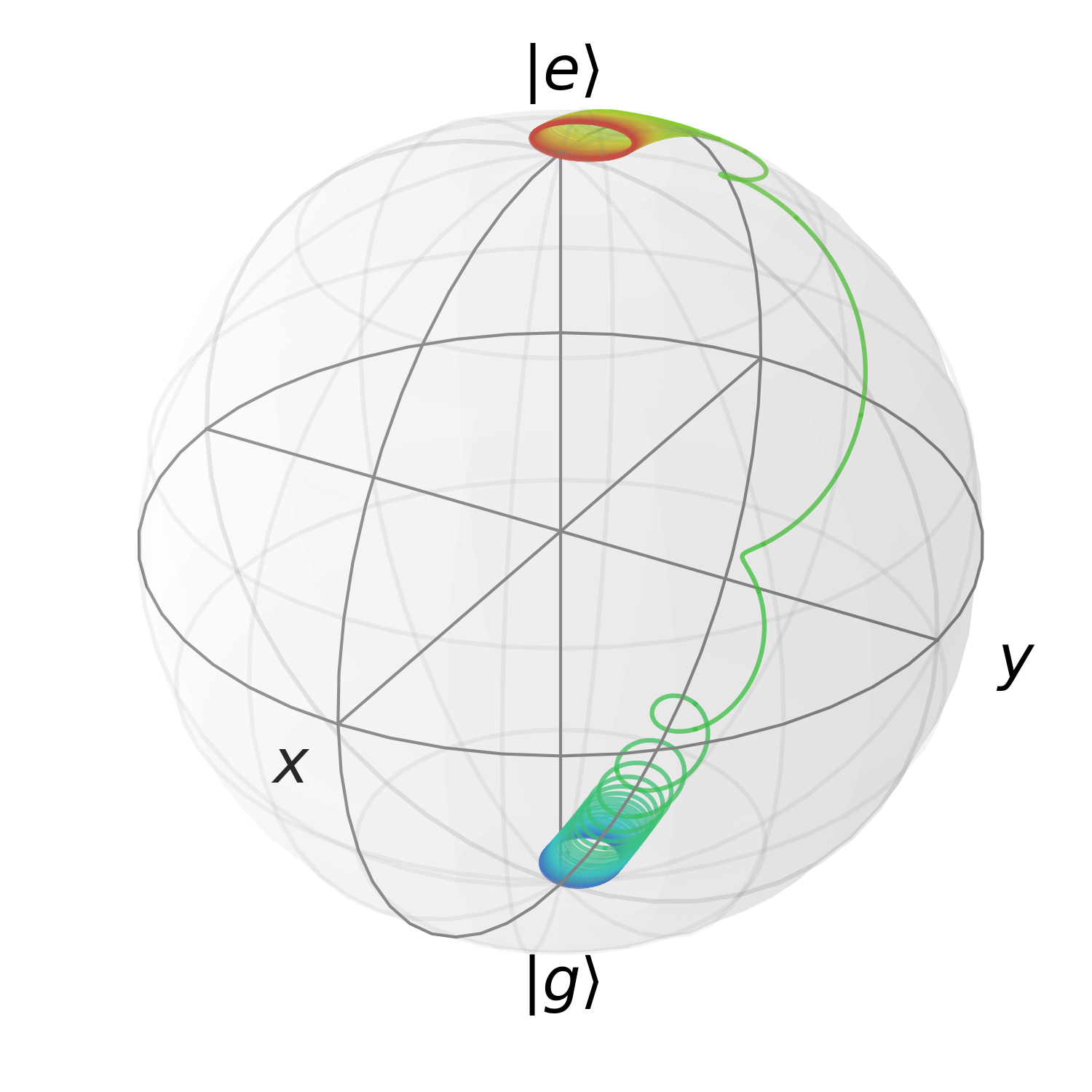}

\end{subfigure}%
\begin{subfigure}{.33\textwidth}
  \centering
  \includegraphics[width=1\linewidth]{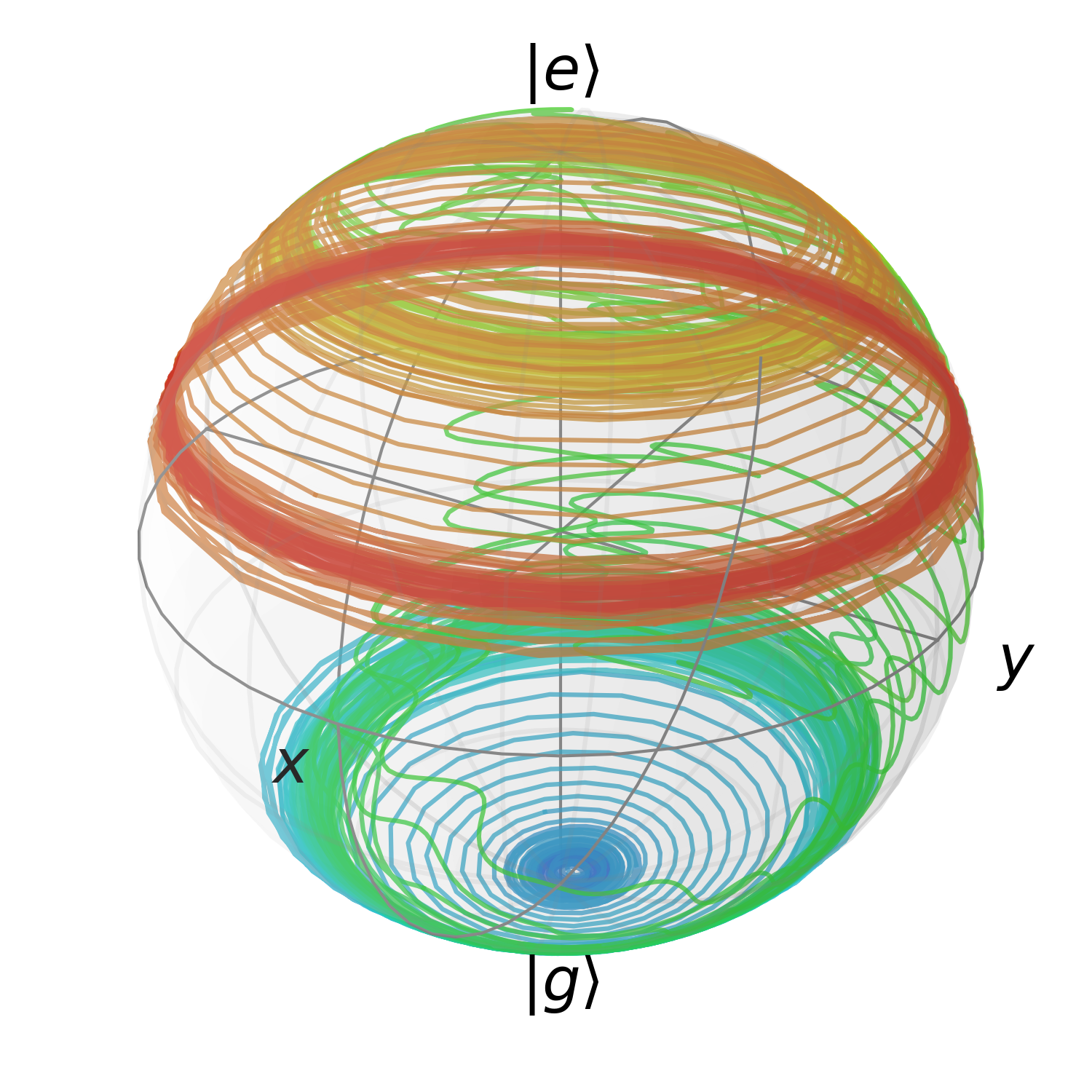}

\end{subfigure}
\begin{subfigure}{.33\textwidth}
  \centering
  \includegraphics[width=1\linewidth]{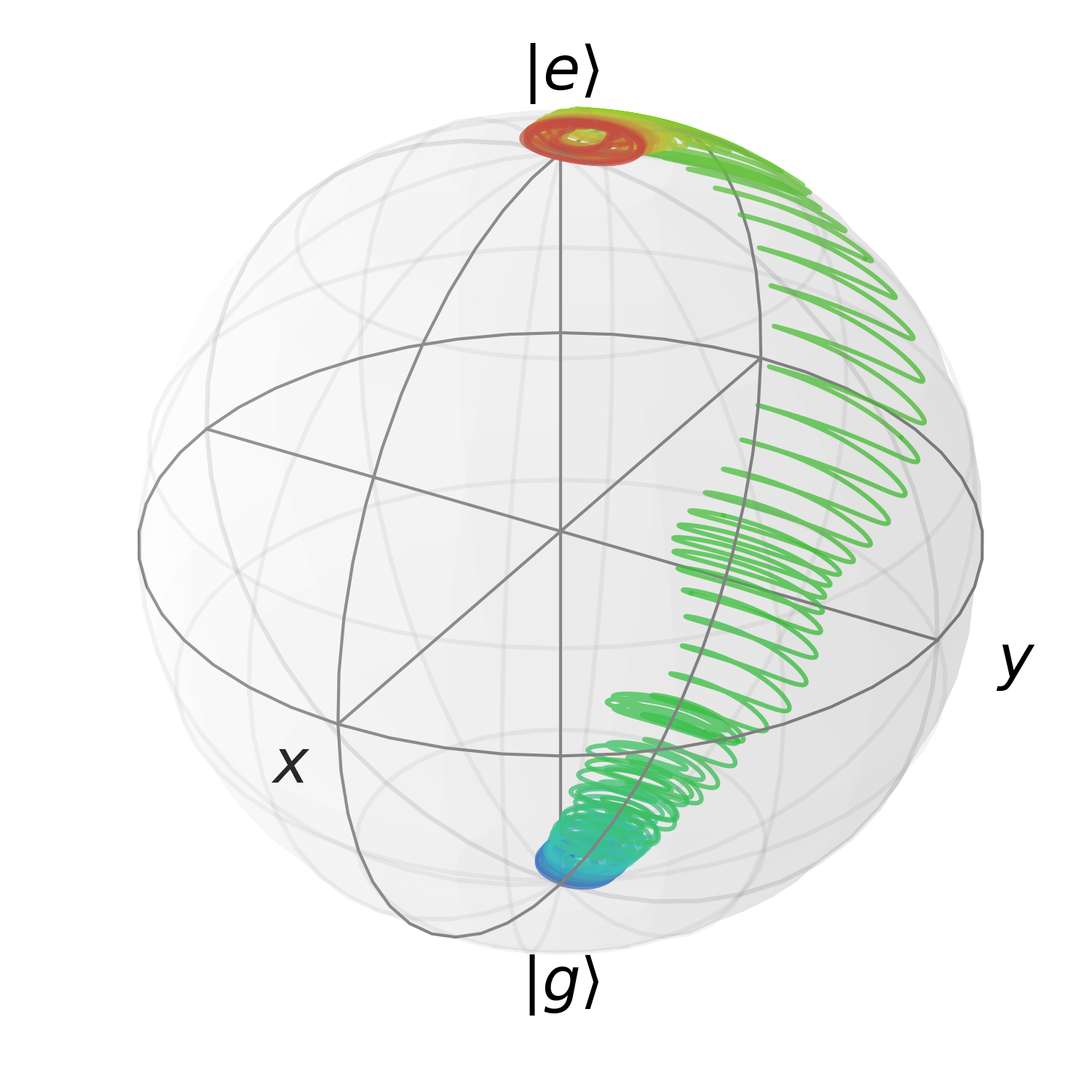}
\end{subfigure}

\begin{subfigure}{.33\textwidth}
  \centering
  \includegraphics[width=1\linewidth]{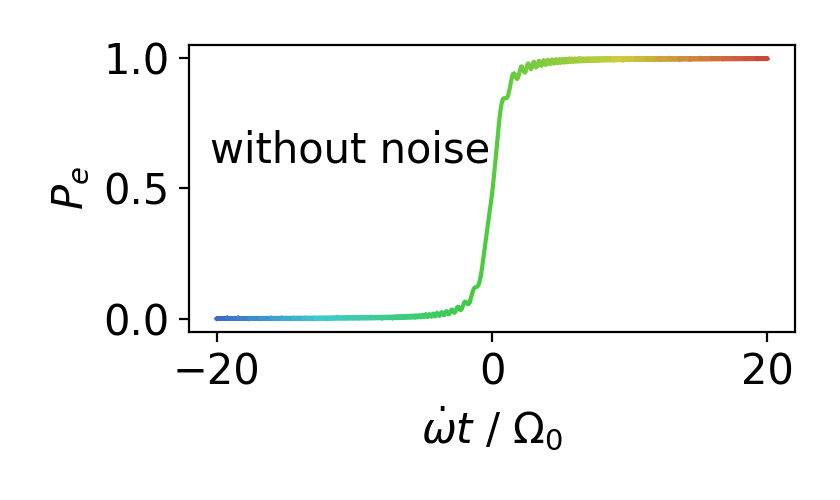}
  \caption{}
  \label{fig:noiseless}
\end{subfigure}%
\begin{subfigure}{.33\textwidth}
  \centering
  \includegraphics[width=1\linewidth]{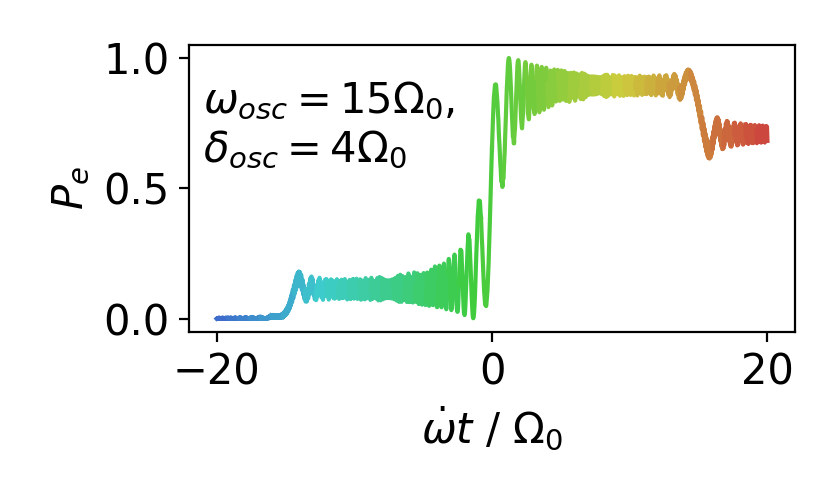}
  \caption{}
  \label{fig:bad example}
\end{subfigure}
\begin{subfigure}{.33\textwidth}
  \centering
  \includegraphics[width=1\linewidth]{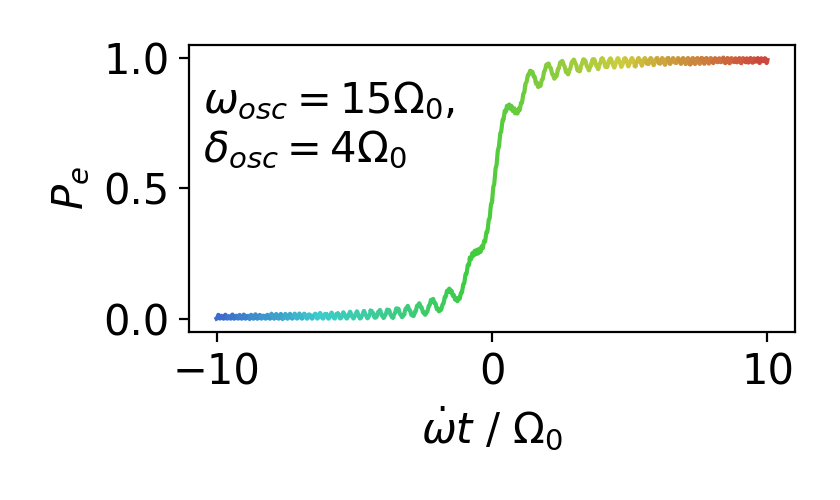}
  \caption{}
  \label{fig:out of range}
\end{subfigure}
\caption{
Bloch sphere trajectory (first row) and excited state population as a function of time (second row) of three example ARP sweeps. {(a) shows an example trajectory for a noiseless ARP sweep; (b) and (c) show two sweeps with identical noise, and the only difference between them is their sweep range. (b) sweeps from -20$\Omega_0$ to $20\Omega_0$, which includes the $15\Omega_0$ noise frequency; whereas (c) sweeps from $-10\Omega_0$ to $10\Omega_0$, excluding the noise frequency. Note the different horizontal scale in (c).}}
\label{fig:bloch spheres}
\end{figure*}

    Without the noise term $\delta(t)$, the Hamiltonian becomes
    \begin{equation}
        \hat{H}(t)= \frac{\Omega_0}{2}\ket{e}\bra{g} + \frac{\Omega_0}{2}\ket{g}\bra{e} - \dot{\omega}t\ket{e}\bra{e}.
    \end{equation}
    Under the assumption that $\omega(t\to-\infty) = -\infty$, $\omega(t\to \infty)=\infty$, the problem has a well-known analytic solution: the fraction of atoms that undergo a diabatic transition is given by the Landau-Zener formula \cite{LZ}
    \begin{equation}
        P_{lost} = \exp \left(-\frac{\pi \Omega_0^2}{2 \dot{\omega}}\right),
    \end{equation}
    which means the fraction of atoms that end up in the exited state is $1 - \exp \left(-\frac{\pi \Omega_0^2}{2 \dot{\omega}}\right)$. In the limit of slow sweeps ($\dot{\omega}\ll \Omega_0^2$), all atoms adiabatically follow the dressed state and end up in the excited state. An example of a noiseless ARP sweep over a finite frequency range ($\omega = \omega_0-20\Omega_0$ to $\omega = \omega_0 + 20\Omega_0$) is presented in Figure \ref{fig:noiseless}, which shows the trajectory of the Bloch vector on the Bloch sphere, and the excited state population over time. The atomic population that starts in the ground state makes a transition as detuning sweeps across zero, and ends up in the excited state by the end of the process.
    
\medskip 

    With a sinusoidal perturbation $\delta(t)=\delta_{osc}\cos(\omega_{osc}t+\phi)$, the evolution becomes more complicated primarily because additional resonances occur when $\dot{\omega}t$ reaches an integer multiple of the noise frequency $\omega_{osc}$ \cite{ashhab}, as exemplified in Figure \ref{fig:bad example}. {The occurrence of these additional resonances can be understood as the result of frequency modulation on the driving field. The frequency modulation leads to sidebands with frequencies $m \omega_{osc}$ ($m = 0, \pm 1, \dots$) away from the carrier frequency. The sidebands resonate with the system during the sweep when the aforementioned condition is reached, giving rise to additional ``noise resonances."} 
    Such noise resonances are problematic in an ARP process, because they can compromise the efficiency of population transfer. As the detuning sweeps across these noise resonances, unwanted population transfer occurs, resulting in a suboptimal excited state population at the end of the process. This becomes clear in the example shown in Figure \ref{fig:bad example}. In a noisy ARP sweep, beside the population transfer near zero detuning, there are two other population jumps (which show up on the Bloch sphere as spirals) that occur when the resonance condition with the noise is reached. As a consequence, the final transfer efficiency is well below the ideal result.

\medskip

To confirm that the noise affects population transfer primarily by means of noise resonances, rather than some other interaction with the system, we present another example in Figure \ref{fig:out of range}, which shows an ARP sweep under the same noise as the previous example, but with a shortened sweep range, so that the noise frequency falls outside the range of detuning chirp, which means the noise resonance condition is never reached. The noise in this case causes the Bloch vector trajectory to oscillate about the adiabatic trajectory in Figure~\ref{fig:noiseless}, but without significant deviations. Accordingly, the transfer efficiency of the sweep is comparable to that of the noiseless case.

\subsection{Noise parameters and ARP efficiency}

While it seems like the efficiency loss associated with the noise can be circumvented by simply shortening the range of the sweep, this is not always possible since the noise in an experiment might have multiple frequency components or even a continuous frequency spectrum. It is therefore worthwhile to investigate the relation between population transfer efficiency of the ARP sweep, $P_e$, and parameters of the noise, namely its frequency $\omega_{osc}$, amplitude $\delta_{osc}$, and phase $\phi$. {We note that while population transfer can have a phase dependence, the goal of our investigation is to understand the effect of noise, where the phase is unknown and varies from shot to shot. Therefore in the following discussion, the results we present are always phase-averaged.}

\medskip

\begin{figure*}[p]
    \centering
    \includegraphics[scale=0.7]{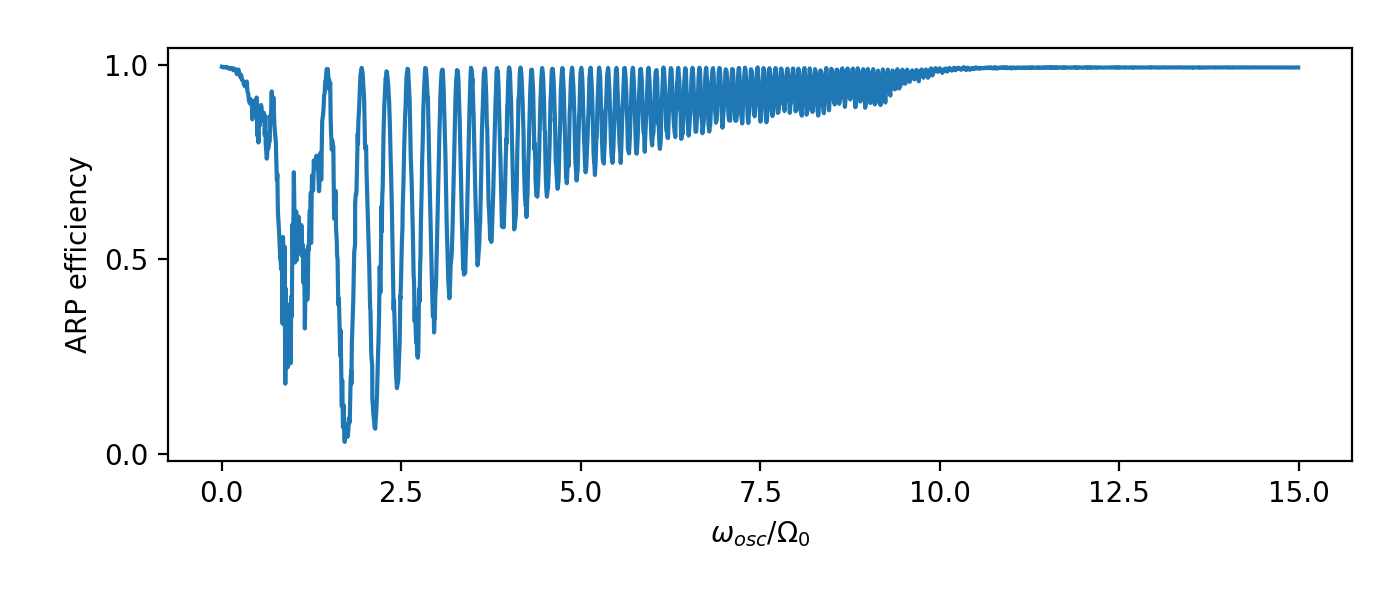}
    \caption{Population transfer efficiency of ARP versus noise frequency. The sweep rate and noise amplitude are $\dot{\omega}=0.2\Omega_0^2$ and $\delta_{osc}=1\Omega_0$. The sweep starts $10\Omega_0$ below resonance and stops $10\Omega_0$ above resonance.}
    \label{fig:pe vs freq}
\end{figure*}

Figure \ref{fig:pe vs freq} shows the efficiency of noisy ARP sweeps with a fixed noise amplitude and variable frequency. Note that the efficiency goes to unity as the noise frequency approaches zero, which corresponds to the limiting case of Fig.~\ref{fig:small freq}. The efficiency also goes to 1 as the noise frequency exceeds the range of the sweep ($10\Omega_0$), as discussed in the previous section. In between these two limits, population transfer depends critically on the noise frequency: in this example, a noise with frequency $1.8 \Omega_0$ leads to almost zero transfer, whereas a noise with frequency $2 \Omega_0$ results in a transfer efficiency of 1. $P_e$ oscillates as a function of $\omega_{osc}$ in this regime. The oscillation becomes faster as $\omega_{osc}$ increases and is modulated by an envelope which goes to 1 in the limits of both small and large $\omega_{osc}$. {The spacing between adjacent peaks of the oscillating function depends inversely on $\omega_{osc}$.}

\begin{figure*}[p]
\centering
\begin{subfigure}{.4\textwidth}
\centering
    \includegraphics[width=1\linewidth]{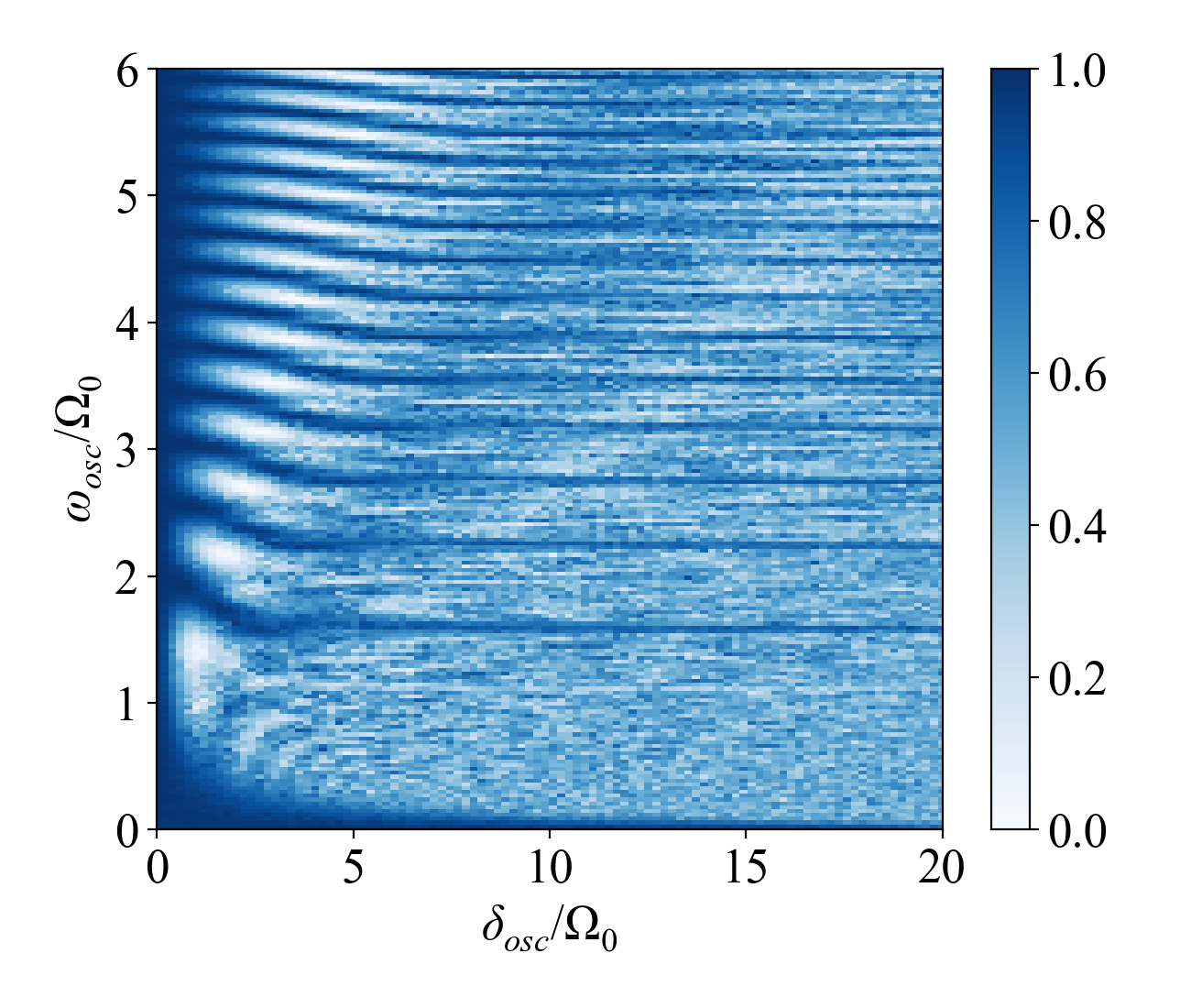}
  \caption{$\dot{\omega}=0.4\Omega_0^2$.}
  \label{fig:simulation low freq}
\end{subfigure}
\begin{subfigure}{.4\textwidth}
  \centering
  \includegraphics[width=1\linewidth]{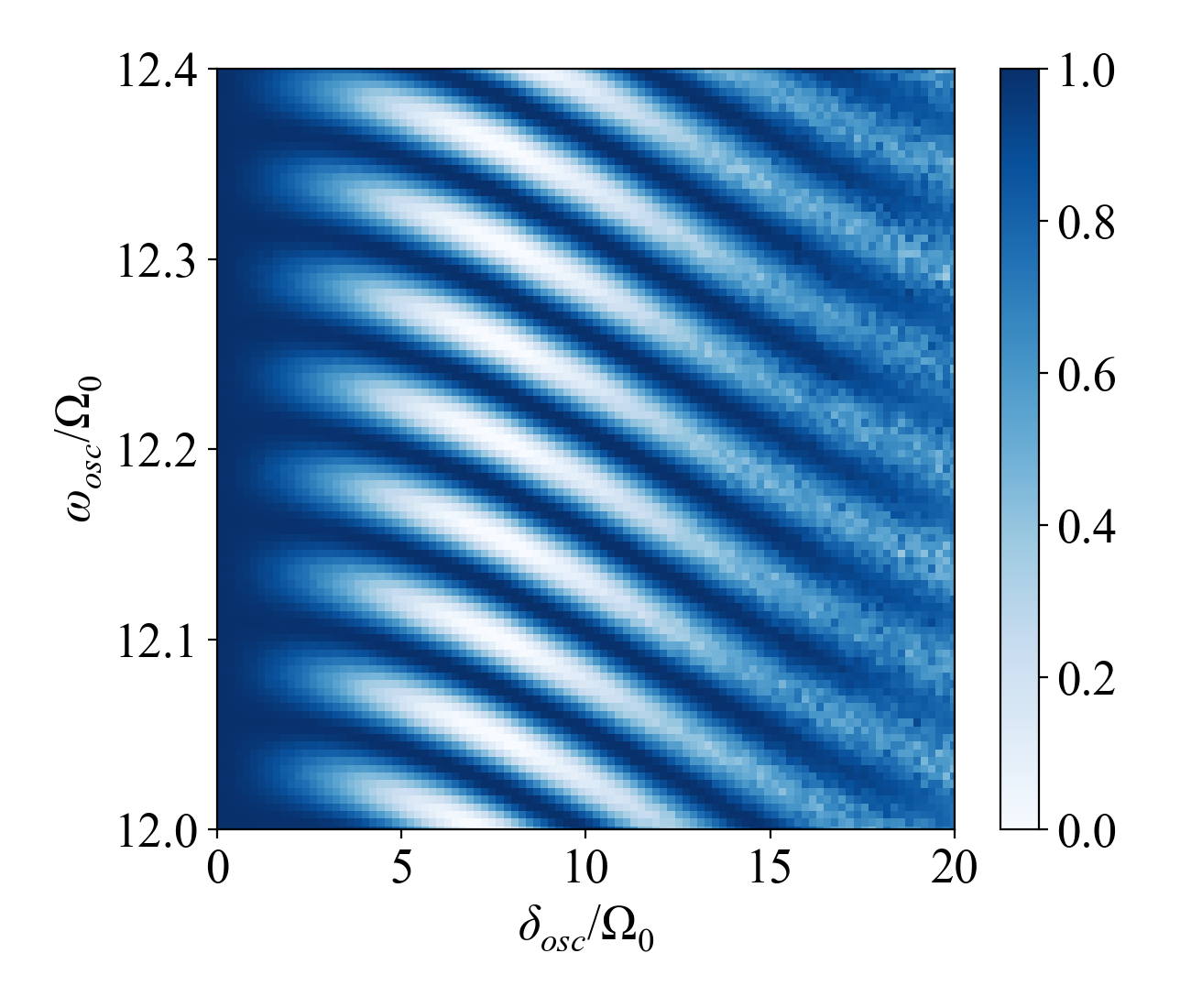}
  \caption{$\dot{\omega}=0.2\Omega_0^2$}
  \label{fig:sub1}
\end{subfigure}%

\begin{subfigure}{.4\textwidth}
  \centering
  \includegraphics[width=1\linewidth]{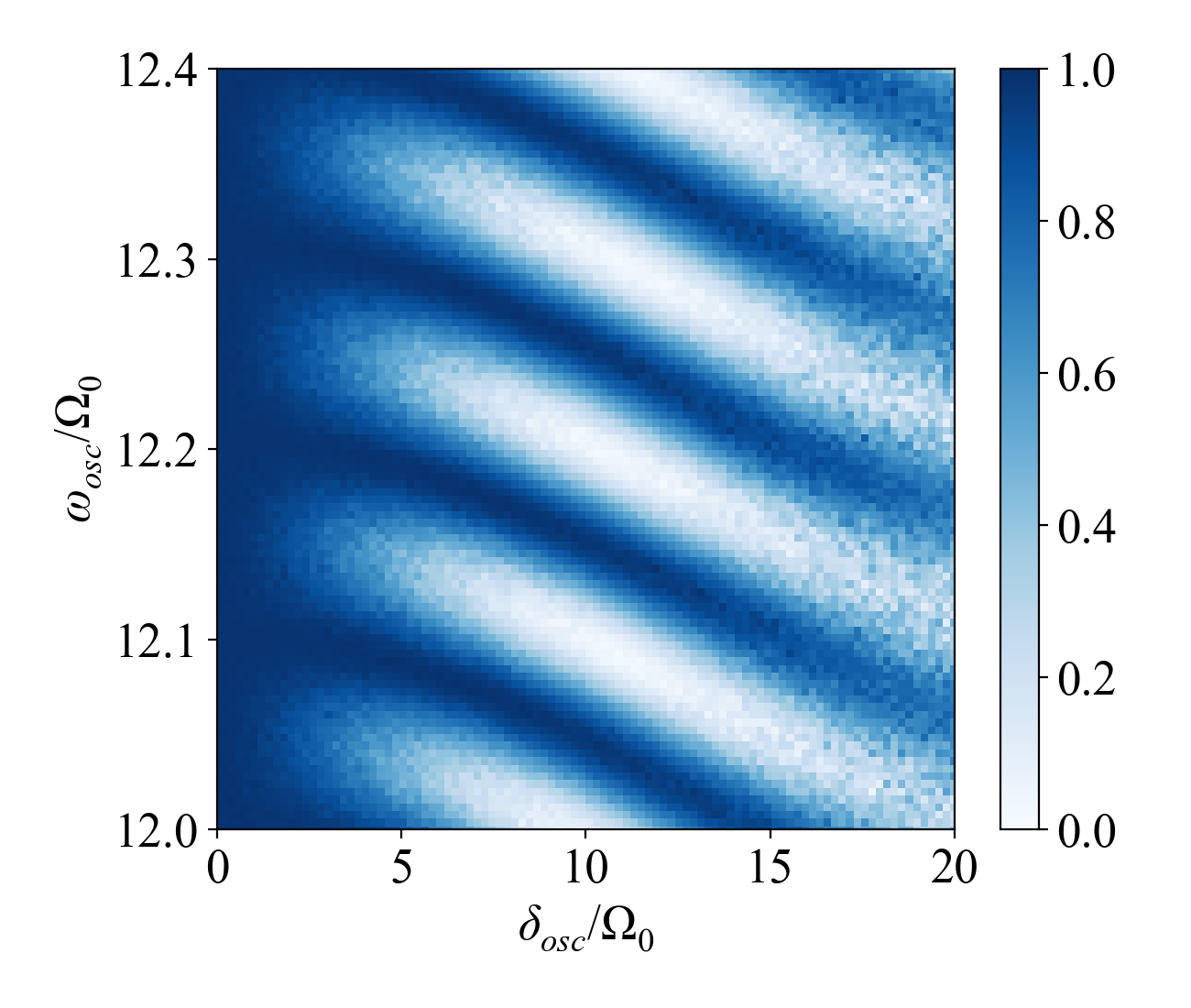}
  \caption{$\dot{\omega}=0.4\Omega_0^2$}
  \label{fig:sub222}
\end{subfigure}

\caption{Population transfer efficiency of ARP versus noise frequency (in the vicinity of $12\Omega_0$) and amplitude, for two different sweep rates. Note that the vertical spacing between adjacent stripes doubles when the sweep rate is doubled.}
\label{fig:color plots}
\end{figure*}

\medskip
Figure \ref{fig:simulation low freq} plots $P_e$ as a function of both $\omega_{osc}$ and $\delta_{osc}$. The graph has a striped pattern, where the spacing between adjacent stripes is consistent with our previous observation, i.e. $\sim$ proportional to $1/\omega_{osc}$. This spacing also depends inversely on the sweep rate, as shown in Figure \ref{fig:sub1} and \ref{fig:sub222}. Note that Figure \ref{fig:sub1} and \ref{fig:sub222} only shows a small range of noise frequency around $12\Omega_0$. Many other notable features of Figure \ref{fig:color plots}, such as the tilting of the stripes, and the vanishing of the strip patterns at large amplitudes, will be further explored in the following section in comparison to a simplified model.

\subsection{Comparison with the ``multi-jump" model}
    The same Hamiltonian is studied in \cite{model}, but in the diabatic regime ($\dot{\omega}\gg \Omega_0^2$), {and in the context of optimal control}. A simplified model (the ``multi-jump" model) for the state's evolution is proposed, which focuses on what happens when the detuning crosses a noise resonance. Every sweep across a resonance is modeled by multiplying the state vector with a unitary matrix. By construction, this model simulates only the population jumps at resonances $\dot{\omega}t = m\omega_{osc}$, and disregards the small oscillations that occur in between.
    
    \medskip
    
    By comparing predictions of the multi-jump model with our simulations, we find that it is a good predictor of the final population transfer in the fast-sweep regime, which is the regime for which the model is intended. But even outside its native regime, the model performs surprisingly well in capturing some important features of the population transfer efficiency $P_e$.
    
    \medskip
    
\begin{figure*}[h]
\centering
\begin{subfigure}{.4\textwidth}
  \centering
  \includegraphics[width=1\linewidth]{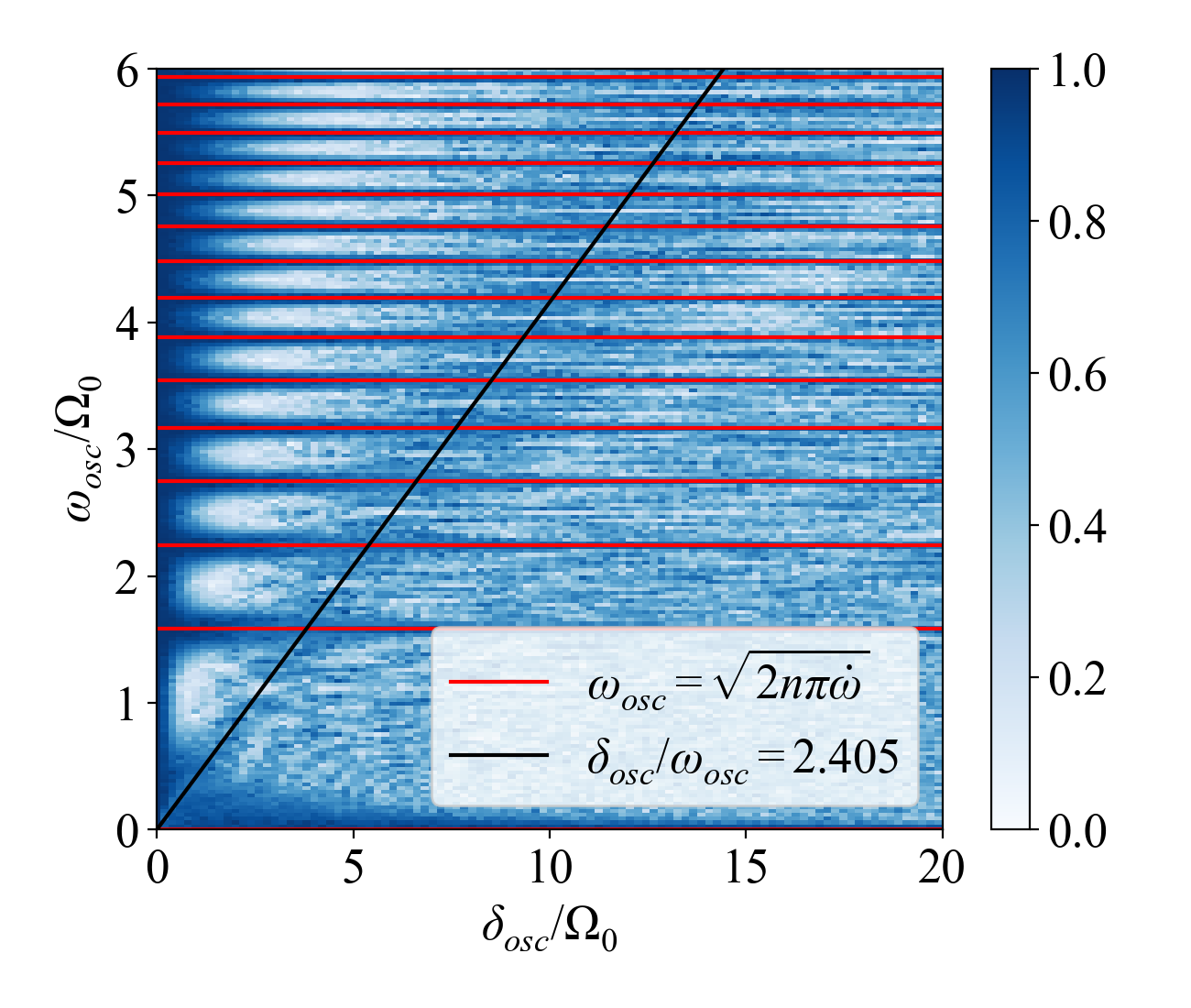}
  \caption{Multi-jump model}
  \label{fig:model low freq with lines}
\end{subfigure}%
\begin{subfigure}{.4\textwidth}
  \centering
  \includegraphics[width=1\linewidth]{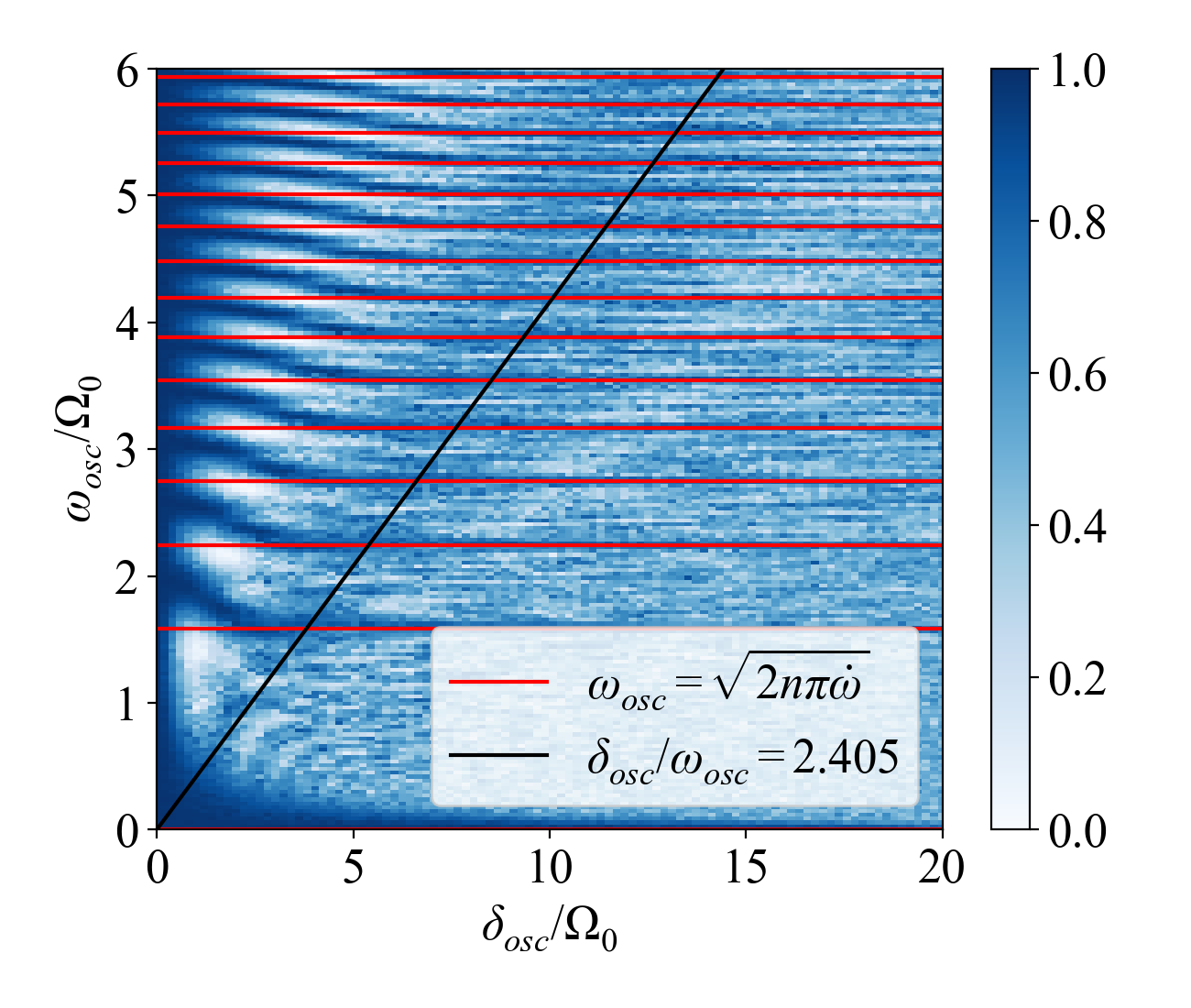}
  \caption{Simulation}
  \label{fig:simulation low freq with lines}
\end{subfigure}
\caption{Population transfer efficiency of ARP predicted by (a) the multi-jump model and (b) by our simulations. For both plots, the sweep rate is $\dot{\omega}=0.4\Omega_0^2$.}
\label{fig:low freq color plots}
\end{figure*}

    Figure \ref{fig:low freq color plots} compares the ARP efficiencies predicted by our simulations and by the multi-jump model. The simplified model makes two important predictions: a horizontal striped pattern that becomes less distinct when $\delta_{osc} \gtrsim \omega_{osc}$; and regions of maximum population transfer when $\omega_{osc}=\sqrt{2n\pi \dot{\omega}}$ ($n = 0, 1, 2, \dots$). These conditions are marked in Figure \ref{fig:model low freq with lines} and \ref{fig:simulation low freq with lines}, which display the prediction by the model and the result of our simulations. While $\omega_{osc}=\sqrt{2n\pi \dot{\omega}}$ does not {predict the correct maxima in the simulations}, it gives the correct vertical spacing between adjacent strips, which, as we observed in the previous section, is inversely related to $\omega_{osc}$ and positively related to sweep rate $\dot{\omega}$. The multi-jump model fails to predict the tilting and distortion of the striped pattern, and we have yet to find an intuitive explanation for this observation.
    
    \medskip
    
\begin{figure}[H]
        \centering
        \begin{subfigure}{0.45\textwidth}
          \centering
          \includegraphics[scale=0.8]{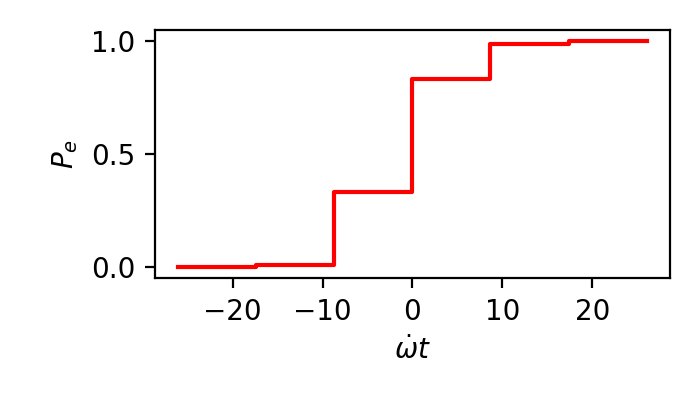}
          \caption{Maximum transfer at $\omega_{osc}=\sqrt{60\pi\dot{\omega}}$}
          \label{fig:jumpMax}
        \end{subfigure}
        
        \vspace{10pt}
        
        \begin{subfigure}{0.45\textwidth}
          \centering
          \includegraphics[scale=0.8]{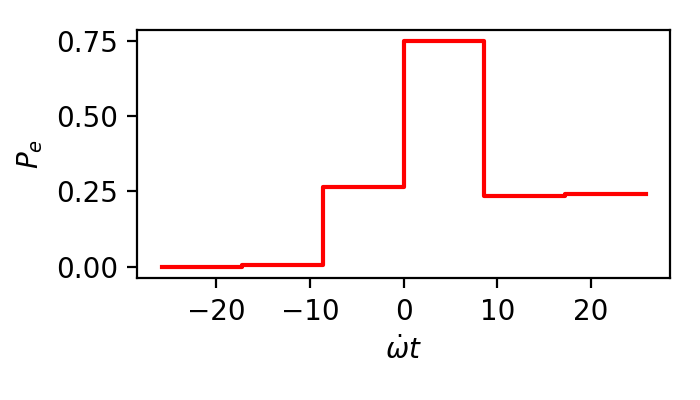}
          \caption{Minimum transfer at $\omega_{osc}=\sqrt{59\pi\dot{\omega}}$}
          \label{fig:jumpMin}
        \end{subfigure}
        \caption{Examples of time-evolution predicted by the multi-jump model, at maximum and minimum transfer conditions. Here, $\dot{\omega}=0.4\Omega_0^2$, $\delta_{osc}=5\Omega_0$.}
        \label{fig:multiJumpEx}
    \end{figure}

    The maximum transfer condition predicted by the multi-jump model is equivalent to requiring that the noise be in phase at all noise resonances ({i.e.~when $\dot{\omega}t=m\omega_{osc}$, where $m = 0, \pm 1, \dots$}). At the point in the ARP sweep when sidebands of the frequency-modulated driving field become resonant with the system, if they all have the same phase, then they contribute constructively to the population transfer (Fig.~\ref{fig:jumpMax}); if the phase at adjacent noise resonances is $\pi$ out of phase, then they contribute destructively (Fig.~\ref{fig:jumpMin}).
   
    \medskip
    The critical ratio $\delta_{osc}/\omega_{osc}=2.405$ is marked in Figure \ref{fig:low freq color plots}. This is roughly the boundary of the striped pattern, and comes from the first zero of the Bessel function $J_0(\delta_{osc}/\omega_{osc})$. According to \cite{model}, this is the condition for the bare resonance ($\dot{\omega}t=0$) to be suppressed by the population transfer at the noise resonances.
    
\subsection{Noise tolerance}
From a practical perspective, it is useful to quantify the level of noise an ARP sweep can tolerate, while still having a reasonably high population transfer efficiency. Since the standard for high efficiency can differ between applications, we proceed by considering an arbitrary threshold $x$ between 0 and 1. We define any ARP sweep with efficiency $P_e > xP_{LZ}$ to be ``acceptable," where $P_{LZ}$ is the ideal efficiency given by the Landau-Zener formula, $P_{LZ}=1-\exp\left(-\frac{\pi \Omega_0^2}{2 \dot{\omega}}\right)$.

\medskip

\begin{figure}[H]
\centering
\begin{subfigure}{.37\textwidth}
  \centering
  \includegraphics[width=1\linewidth]{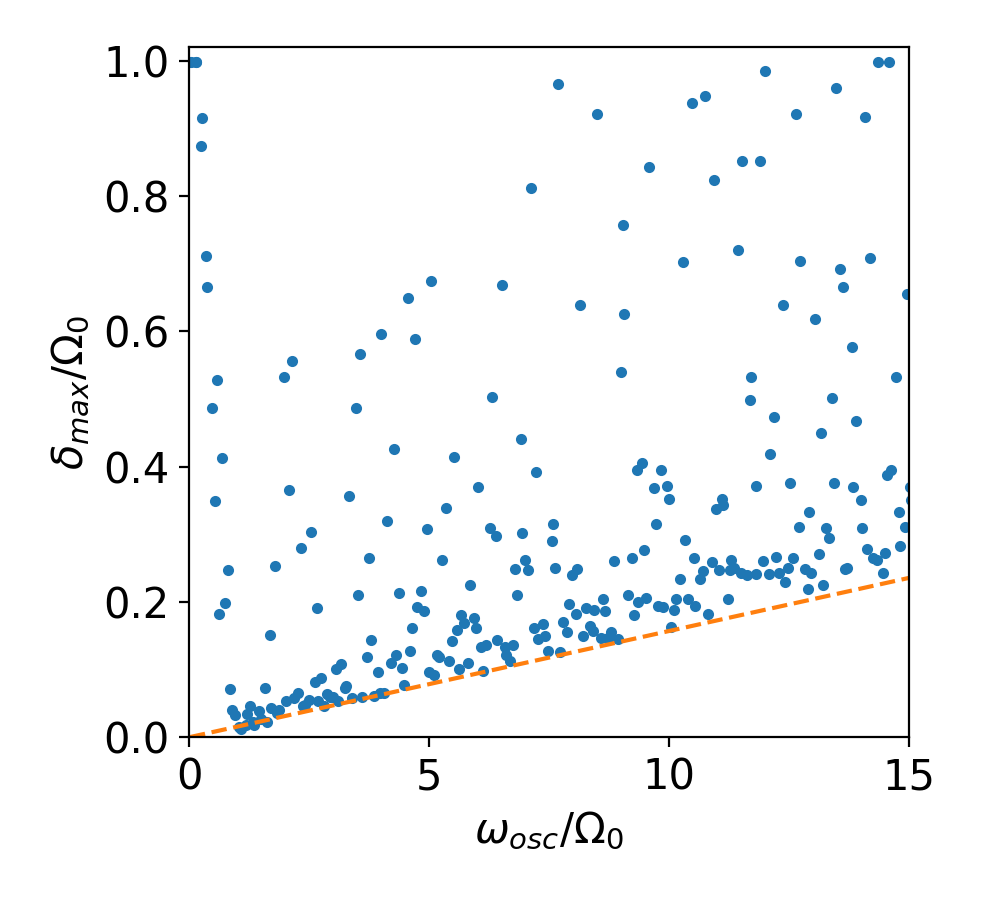}
  \caption{}
  \label{fig:max amp vs freq}
\end{subfigure}%

\begin{subfigure}{.37\textwidth}
  \centering
  \includegraphics[width=1\linewidth]{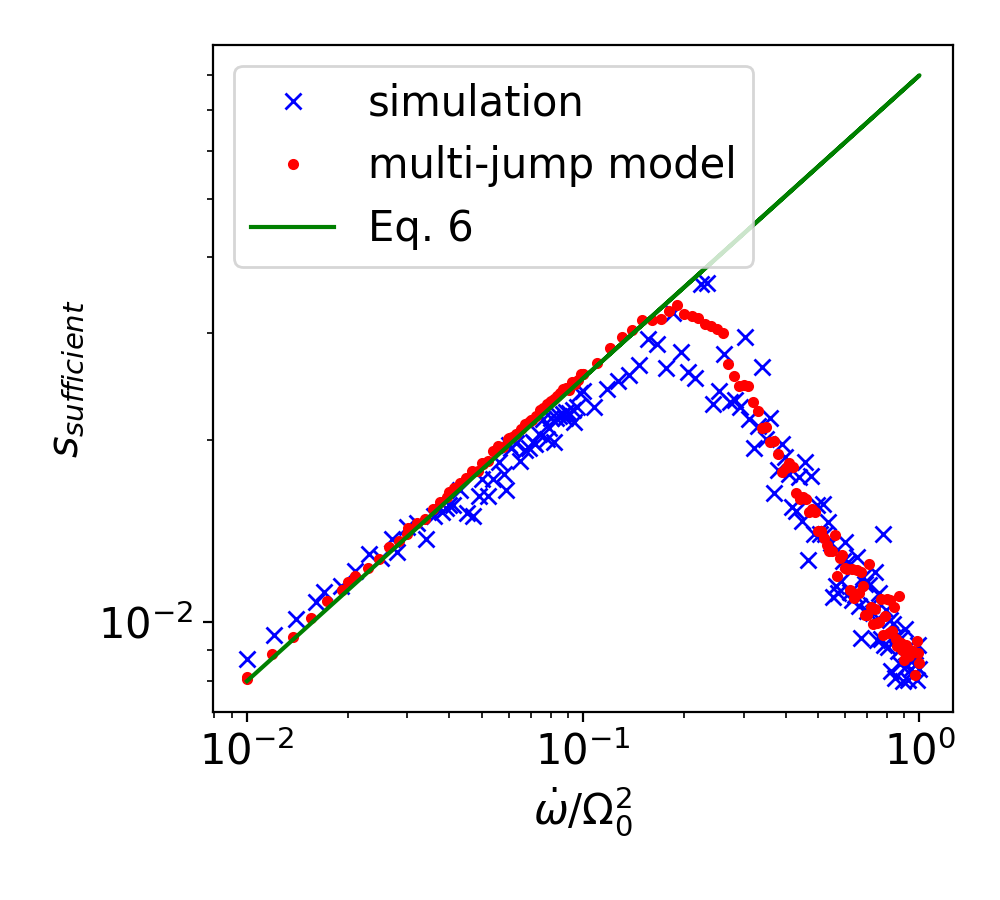}
  \caption{}
  \label{fig:slope vs rate}
\end{subfigure}
\caption{(a) shows an example of the maximum acceptable noise amplitude versus noise frequency, with a threshold value of $x=0.99$, and sweep rate $\dot{\omega}=0.05\Omega_0^2$; (b) plots the slope of the sufficient lower boundary, $s_{\text{sufficient}}$, versus sweep rate $\dot{\omega}$, for threshold $x=0.99$. We see that Eq. \ref{sufficient} gives a good approximation for $s_{\text{sufficient}}$ in the $\dot{\omega}/\Omega_0^2 \ll 1$ limit.}
\label{fig:noise tolerance}
\end{figure}

One way to quantify noise tolerance is to determine, for a given frequency of noise, the maximum amplitude the noise can have such that the efficiency of the sweep is still acceptable. This characterization is done in Figure \ref{fig:max amp vs freq}, which shows the maximum acceptable amplitude for a range of noise frequencies, with a sweep rate of $\dot{\omega}=0.05\Omega_0^2$ and threshold of $x=0.99$. 
We observe that while the maximum allowable noise amplitude varies considerably with small changes in $\omega_{osc}$, a sufficient lower boundary exists and depends linearly on $\omega_{osc}$. Below this sufficient condition (the orange line in Figure \ref{fig:max amp vs freq}), the population transfer is guaranteed to be above the threshold value (in this example, 99\%).

Figure \ref{fig:slope vs rate} demonstrates that the slope of this sufficient boundary depends on the sweep rate of the ARP, where the acceptable threshold is set to $x=0.99$. Here, we notice that the slope($\dot{\omega}$) appears linear (on a logarithmic scale), up to $\dot{\omega}/\Omega_0^2 \sim 0.1$. The same is true for results by the multi-jump model, which are plotted in red. 

\medskip 
Based on the multi-jump model, keeping only the contribution of -1st, 0th, and 1st order resonances {(which is a good approximation when the noise is not too strong, because the higher order sidebands due to frequency modulation are negligible when modulation amplitude is small)}, we find an explicit expression for the excited state fraction $P_e$, which depends on $\dot{\omega}, \omega_{osc}, \delta_{osc} $ and $\phi$. From here we obtain a lower bound for $P_e$, assuming $\dot{\omega}\ll \Omega_0^2$ and $\delta_{osc} \ll \omega_{osc}$, given by
\begin{equation}
    P_{min} = \left(1-\exp\left(-\frac{\pi \Omega_0^2}{2\dot{\omega}} \right)\right)\left(1-2\exp \left(-\frac{\pi \delta_{osc}^2}{8\dot{\omega}\omega_{osc}^2} \right) \right)^2
\end{equation}
Then a sufficient condition for $P_e > xP_{LZ}$ is $P_{min} > xP_{LZ}$, which leads to 
\begin{equation}
    \left(1-2\exp \left(-\frac{\pi \delta_{osc}^2}{8\dot{\omega}\omega_{osc}^2} \right) \right)^2 > x
\end{equation}
and thus
\begin{equation}
   \frac{\delta_{osc}}{\omega_{osc}} < s_{\text{sufficient}} \coloneqq \sqrt{-\frac{8}{\pi}\ln(\frac{\sqrt{x}+1}{2})}\sqrt{\frac{\dot{\omega}}{\Omega_0^2}} 
   \label{sufficient}
\end{equation}
Eq.~\ref{sufficient} provides a sufficient condition for an ARP efficiency to be above the threshold $x$. As shown in Figure \ref{fig:slope vs rate}, up to $\dot{\omega}/\Omega_0^2\sim 0.1$, Eq.~\ref{sufficient} agrees excellently with the multi-jump points, and is close to the simulation points. The discrepancy is due to the fact that the model is used outside of its intended regime (the diabatic limit).

\section{Conclusion}
By performing numerical simulations, we find that, in the presence of sinusoidal noise, the efficiency of ARP is influenced by resonances that occur between the system and the noise. The noise has less of an effect, in general, if the amplitude is small, or if its frequency is low or high. Specifically, if the noise frequency exceeds the full range over which the detuning is chirped, then its effect on population transfer is almost negligible. The ARP efficiency depends critically on noise frequency in the intermediate frequency regime. The regions in parameter space with maximum transfer efficiency are similar to those predicted by the diabatic multi-jump model. While the functional form of the efficiency is complicated, we propose a sufficient condition to achieve high population transfer by studying the worst-case scenario, which can be used to quantify noise tolerance in an ARP process.

\section{Acknowledgements}

The authors would like to thank Joseph McGowan and Nick Mantella for helpful discussions, as well as Dr.~Yuval Sanders for kindly answering our questions about his paper \cite{model}. This work was supported by the Natural Sciences and Engineering Research Council of Canada. AMS is a Fellow of CIFAR.

\nocite{*}

\bibliography{references}
\bibliographystyle{apsrev4-1}
\end{document}